\documentclass[preprint,aps,amsmath,amssymb,pre]{revtex4}

\usepackage{subfigure}
\usepackage{graphicx}

\newcommand{\bc}{\begin{center}}
\newcommand{\ec}{\end{center}}
\newcommand{\be}{\begin{equation}}
\newcommand{\ee}{\end{equation}}
\newcommand{\ba}{\begin{array}}
\newcommand{\ea}{\end{array}}
\newcommand{\beq}{\begin{eqnarray}}
\newcommand{\eeq}{\end{eqnarray}}
\newcommand{\xx}{{\bf x}}
\newcommand{\hh}{{\cal H}}

\begin{document}

\title{Path integral Monte Carlo study of the interacting quantum double-well
model: quantum phase transition and phase diagram}

\author{
Dong-Hee~Kim
}
\affiliation{
Department of Physics, Korea Advanced Institute of Science and Technology, Daejeon 305-701, Republic of Korea}
\author{Yu-Cheng~Lin}
%\email[Corresponding author: ]{lin@lusi.uni-sb.de} 
\author{Heiko~Rieger}
\affiliation{
Theoretische Physik, Universit\"at des Saarlandes, 66041 Saarbr\"ucken, Germany}

\begin{abstract}
The discrete time path integral Monte Carlo (PIMC) with a one-particle
density matrix approximation is applied to study the quantum phase
transition in the coupled double-well chain. To improve the convergence 
properties, the exact action for a single particle in a double well potential is
used to construct the many-particle action. The algorithm is applied to the
interacting quantum double-well chain for which the zero-temperature phase
diagram is determined. The quantum phase transition is studied via finite-size 
scaling and the critical exponents are shown to be compatible with
the classical two-dimensional (2D) Ising universality class -- 
not only in the order-disorder limit (deep potential wells) but also
in the displacive regime (shallow potential wells).

\end{abstract}
\maketitle

\section{Introduction}

\begin{figure}
 {\par\centering \resizebox*{0.9\textwidth}{!}
   {\includegraphics{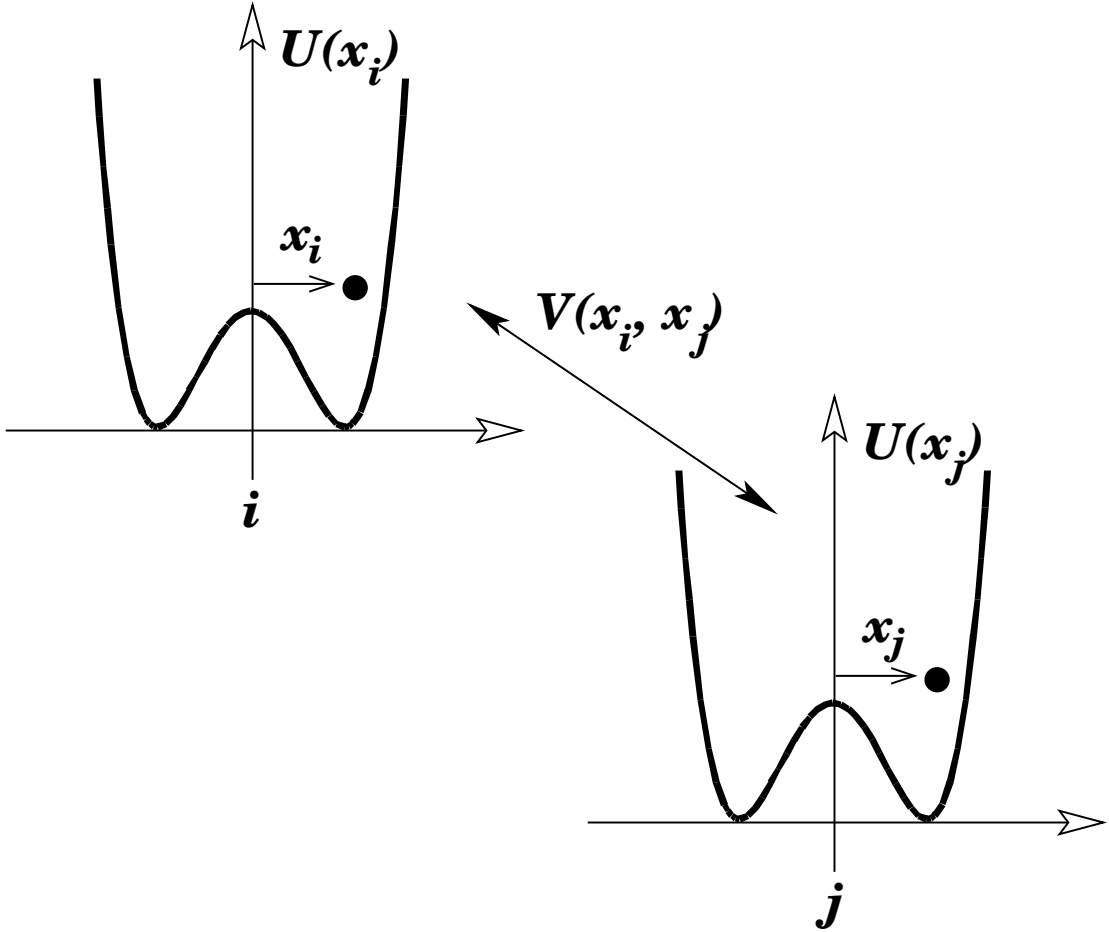}} \par}
 \caption{
 \label{fig:dw}
Schematic representation of a coupled tunneling model in which the local potential
is described by a double-well form.
}
\end{figure}

The two-level tunneling model provides a phenomenological description
of the low-temperature properties of glassy materials\cite{TM}.  In the
simplest case, an isolated tunneling system can be represented by a
particle moving in a double-well potential.  Experimental findings
have suggested that the interactions between the tunneling systems play
a crucial role in the low temperature behavior which deviates from the
predictions of the non-interacting two-level systems\cite{EXP}. The
model Hamiltonian of the system with $L$ particles is then given by
\be
\mathcal{H}=\sum_{i=1}^L \Bigl(\frac{p_i^2}{2\mu}+U(x_i)\Bigr)
+\sum_{j<i} V(x_i,x_j)\;,
\label{ham}
\ee
where $x_i$ is the (one-dimensional) displacement of the $i$-th
particle ($i=1,\ldots,L$) of mass $\mu$ from a reference position,
$p_i=\frac{\hbar}{i}\frac{\partial}{\partial x_i}$ denotes the momentum operator, 
$U(x_i)$ is a local potential for the displacement of the $i$-th particle that is
usually assumed to be a double well potential, and $V(x_i,x_j)$
describes the interaction between particles, see Fig.~\ref{fig:dw}. Apart
from glassy materials, this coupled double-well model has been applied
to other systems, including structural phase transitions of a
wide range of systems, e.g. uniaxial
ferroelectrics\cite{BRUCE}. Most numerical computations devoted
to understanding the interacting double-well model have mainly treated
the problem in the framework of the classical $\phi^4$ model or have
been limited in the "two-state" limit by studying the corresponding
Ising model. These simplifications reveal the difficulties inherent in
simulations of the quantum coupled double-well model.

In this paper we present an efficient path integral Monte Carlo (PIMC)
algorithm to study interacting particles, each of which is confined
to a double well potential. The method is presented in the next section,
and it is applied to the one-dimensional interacting double-well model in the
Sec.~\ref{sec:results}, which also contains the results:
the phase diagram and the discussion of the universality class of the
quantum phase transition.

\section{The method}
\label{sec:methos}

The partition function of (\ref{ham}) is given by
\be
{\cal Z }=\int d\xx\,\rho(\xx,\xx;\beta)\;,
\ee
where $\xx=(x_1,\ldots,x_L)$ is the displacement configuration of the
whole system and
\be
\rho(\xx,\xx';\beta)=\langle\xx|e^{-\beta\hh}|\xx'\rangle
\ee
is the density matrix, with $\beta=1/T$ the inverse
temperature. Observables that are diagonal in the displacements, like
the the $m$-th moment $\langle x_i^m\rangle$, are given by
\be
\langle{\cal O}(\xx)\rangle=\frac{1}{\cal Z}
\int d\xx\,\rho(\xx,\xx;\beta){\cal O}(\xx)\;.
\ee
By splitting the Hamiltonian (\ref{ham}) into two (non-commuting)
parts $\hh=\hh_A+\hh_B$  and using the Suzuki-Trotter identity, 
one arrives at the path integral formula for
$\rho$: 
\be
e^{-\beta(\hh_A+\hh_B)}=
\lim_{M\to\infty} \Bigl[e^{-\tau\hh_A}e^{-\tau\hh_B}\Bigr]^M
\ee
with $\tau=\beta/M$. The conventional choice for $\mathcal{H}_A$ and $\mathcal{H}_B$ is
the kinetic energy for $\mathcal{H}_A$ (which is diagonal in the momenta $p_i$)
and the potential plus interaction energy $U+V$ for $\hh_B$ (which is
diagonal in the displacement variables). In the lowest-order of
the commutator expansion -- the so-called primitive approximation, 
the high-temperature  density matrix  becomes
\be
\rho(\mathbf{x},\mathbf{x'};\tau)=
\prod_{i,j}^L e^{-\frac{\tau}{2}\left(U(x_i)+V(x_i,x_j)\right)}
   \rho_0(x_i,x'_i;\tau) e^{-\frac{\tau}{2}\left(U(x'_i)+V(x'_i,x'_j)\right)}
\ee
where $\rho_0(x_i,x'_i;\tau)$ is the free particle density matrix. This choice leads 
to bad convergence properties in the Trotter number $M$ \cite{CEPERLEY}
because of the fractal character of a trajectory of a free quantum
mechanical particle described by the term $\hh_A$. 

The purpose of this paper is to demonstrate the efficiency of another
choice for $\hh_A$ and $\hh_B$ by treating the single particle diffusion 
within a double well potential exactly and separately from the particle
interactions. Doing this, we have
\beq
\hh_A&=&\sum_{i=1}^L \frac{p_i^2}{2\mu}+U(x_i)\;\nonumber\\
\hh_B&=&\sum_{i<j} V(x_i,x_j)\;.
\label{hamab}
\eeq
This strategy is expected to be most promising in the case when the interactions 
are much weaker than the mean potential energy of the particle.

With (\ref{hamab}) the path integral expression for the density matrix
becomes
\be
\rho(\xx,\xx;\beta)=
\lim_{M\to\infty} \int d\xx_1\ldots d\xx_{M-1}
\,\prod_{m=0}^{M-1}
\rho_A(\xx_m,\xx_{m+1};\tau)\rho_B(\xx_m,\xx_{m+1};\tau)
\label{pi}
\ee
with $\xx=\xx_0=\xx_M$ and 
\beq
\rho_A(\xx,\xx';\tau)&=&\prod_{i=1}^L \rho^{(1)}(x_i,x_i';\tau)\nonumber\\
\rho_B(\xx,\xx';\tau)&=&
\prod_{i<j} e^{-\frac{\tau}{2}\left(V(x_i,x_j)+V(x_i',x_j')\right)}\;,
\label{eq:AB}
\eeq
where 
\be
\rho^{(1)}(x,x';\tau)
=\langle x | e^{-\tau\frac{p^2}{2\mu}} | x' \rangle\, e^{-\tau \frac{U(x)+U(x')}{2}}
\label{eq:symm}
\ee
is the one-particle density matrix for a single particle in a
potential $U$. For a double well potential this is not known
analytically but can easily be computed numerically with the 
matrix multiplication method\cite{NMM}. This method is based on the
recursion formula
\be
\rho^{(1)}(x, x';2\delta\tau)=
\int_{x_{\textrm{min}}}^{x_{\textrm{max}}} dx'' 
\rho^{(1)}(x,x'';\delta\tau) \rho^{(1)}(x'',x';\delta\tau)
\ee
and the fact  that in the limit $\delta\tau\to0$ the one-particle
density matrix $\rho^{(1)}$ can be factorized into the kinetic and 
potential energy part:
\be
\rho^{(1)}(x,x'';\delta\tau)
\to
\left(\frac{\mu}{2\pi\hbar^2\delta\tau}\right)^{1/2}
e^{-\frac{\mu(x-x')^2}{2\hbar^2\delta\tau}}
\cdot e^{-\frac{\delta\tau}{2}\left(U(x)+U(x')\right)}.
\ee
By squaring the density matrix $k$ times, we will lower the
temperature by a factor of $2^k$ and reach the required temperature
$\tau$. 
For a given potential $U(x)$, the limits of
integration, $x_{\textrm{min}}$ and $x_{\textrm{max}}$, are chosen appropriately --
not too large for computational reasons and not too small for numerical accuracy.  
Once the limits are set, a fine grid between $x_{\textrm{min}}$ and $x_{\textrm{max}}$ 
should be constructed for the numerical integrations. The spacing between
successive grid points should be sufficiently small to ensure the high
accuracy.  We store this one-particle density matrix in a
two-dimensional array as a look-up table for use during the
simulations, and employ a simple bilinear interpolation to
determine the matrix elements for any point $(x_i, x_i')$ within
$[x_{\textrm{min}}, x_{\textrm{max}}]$ in the continuous position
space.  We note that the symmetric break up of the propagator
in the form of Eq.~(\ref{eq:symm}) satisfies a unitarity condition
$\rho^{(1)}(\delta\tau)\rho^{(1)}(-\delta\tau)=1$, which can be utilized 
to reduce errors resulting from discretization of $\tau$, as discussed
in \cite{UNITARITY}.

Path integral Monte Carlo means the evaluation of the integral
(\ref{pi}) via importance sampling of the configurations
$(\xx_1,\ldots,\xx_{M-1})$ (for fixed $M$) with the appropriate weight
given by the integrand of (\ref{pi}). Here we use a single step update
scheme: Let ${\bf X}=(\xx_1,\ldots,\xx_{M-1})=\{x_{i,m}\}$ be the
current configuration. We generate a new configuration ${\bf X'}$
which differs from the old configuration ${\bf X}$ only in a
single particle displacement in a particular time slice:
$x_{i,m}'=x_{i,m}+\delta$, where $\delta\in[-\varepsilon,\varepsilon]$
is a uniformly distributed random number, and $\varepsilon$ the step
size. The acceptance probability $w({\bf X}\to{\bf X'})$ of this new
configuration should be chosen to fulfill detailed balance with respect to 
the weights of of the old and new configurations; a possible choice is
\beq
w({\bf X}\to{\bf X'})
&=&
\min\left[1,\;
\prod_{m=0}^{M-1}
\frac{\rho_A(\xx_{m-1},\xx_{m}';\tau)\rho_B(\xx_m',\xx_{m+1};\tau)}
{\rho_A(\xx_{m-1},\xx_{m};\tau)\rho_B(\xx_m,\xx_{m+1};\tau)}\right]\nonumber\\
&=&
\min\left[1,\;
e^{-\tau\Delta V({\bf X},{\bf X'})}
\cdot
\frac{\rho^{(1)}(x_{i,m-1},x_{i,m}';\tau)\rho^{(1)}(x_{i,m}',x_{i,m+1};\tau)}
{\rho^{(1)}(x_{i,m-1},x_{i,m};\tau)\rho^{(1)}(x_{i,m},x_{i,m+1};\tau)}\right]\;,
\eeq

\section{The one-dimensional model and results}
\label{sec:results}

\begin{figure}
 {\par\centering \resizebox*{0.9\textwidth}{!}
   {\includegraphics{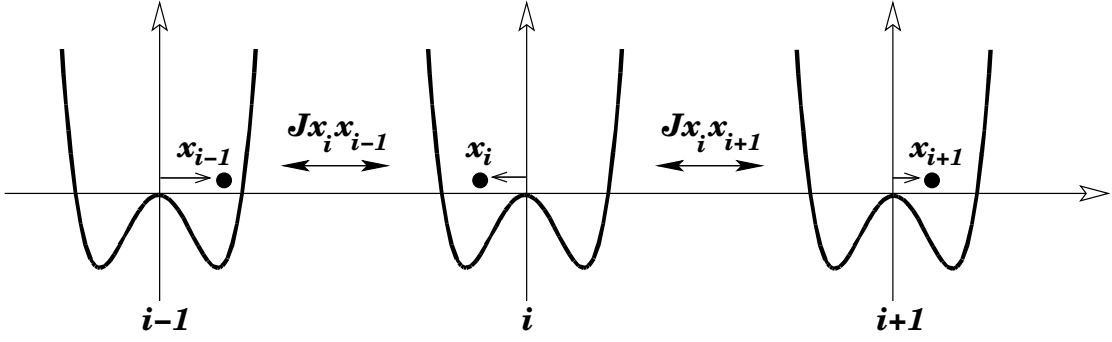}} \par}
 \caption{
 \label{fig:model}
Representation of the model defined in Eq.~(\ref{ham}) in the one-dimensional
form.
}
\end{figure}

To test the above algorithm we focus here on a one-dimensional geometry
in which particles interact only with their nearest neighbors, and
assume $V(x_i,x_{i+1})$ to be quadratic in the displacements 
\be
\hh_B=\sum_{i<j}V(x_i,x_j)=-\sum_{i=1}^L J_i x_i x_{i+1}\;,
\label{VV}
\ee
(cf.\ Fig.~\ref{fig:model}).  Furthermore we choose homogeneous 
interaction strength $J_i=J>0$,  
and the double-well potential in the symmetrical form with two minima located
at $\pm 1$:
\be
U(x)=V_{0}(x^{4}-2x^{2}).
\label{UU}
\ee 
Periodic boundary conditions are imposed.

The model (\ref{ham}) with (\ref{VV}) and (\ref{UU}) has a $Z_2$-symmetry
($x_i\to-x_i\forall i$) and corresponds to a quantum version of a
$\phi^4$ theory, which is expected to belong to the universality class of 
$1+1$-dimensional Ising model.  Suppose that the height of
the potential barrier $V_0$ is large compared to the energy scale of
the particle executing small oscillations in one of the double wells.
The model is then equivalent to the one-dimensional Ising model in a transverse field   
\be
\mathcal{H}_{\textrm{TIM}}=
-\Gamma\sum_i \sigma^x_i-J\sum_{\langle i,j\rangle} \sigma^z_i \sigma^z_j\;,
\ee
where the transverse field $\Gamma$ corresponds to the tunneling splitting in 
the double-well problem. Therefore we expect it to display a zero-temperature 
quantum phase transition (\cite{SACHDEV}) from a disordered phase with
$\langle x_i\rangle=0$ to an ordered phase with $\langle
x_i\rangle\ne0$ at a critical interaction strength $J_c$ (for fixed
$\mu$ and $V_0$). According to the universality hypothesis, the same
universality class, i.e. the 2D Ising, should extent to the region
where $V_0$ is small compared to the interactions between particles,
the so-called displacive region\cite{BRUCE,SCHNEIDER,BINDER}.

For a given value of the parameters $V_0$ and $J$ we computed the following quantities: 
the average of the displacement $m$ (i.e. the magnetization in the spin formulation), 
defined as 
$m(L,M)=\frac{1}{LM}\sum_i^L \sum_n^M  \langle x_{i,n}  \rangle \,$,
where $x_{i,n}$ is the position of the $i$-th particle at the time step $n$ with respect to the zero
position of the local potential $V_{\textrm{dw}}$; 
the fourth-order cumulant of the magnetization given by 
$g=\frac{1}{2} \left ( 3-\left \langle m^4 \right \rangle/\left \langle m^2 \right \rangle^2 \right )$,     
where $\langle \cdots \rangle$ denotes the expectation value over MC configurations; the susceptibility
defined as
$\chi=L\beta \left ( \left \langle m^2 \right \rangle- \left \langle |m|^2 \right \rangle \right)$.
Close to a quantum critical point, one expects\cite{SACHDEV} observables $O$ to scale as
\be
O =  L^{-x_{O}} \tilde{O}\left(\delta\,L^{1/\nu},\frac{\beta}{L^z}\right)
\label{eqn:scaling}
\ee 
where $x_{O}$ is the scaling dimension of the observable $O$, $\nu$  the 
correlation length exponent and $z$  the dynamical exponent. If the transition falls into
the Ising universality class, the dynamical exponent $z$ is unity\cite{SACHDEV}. In the following we assume 
this to be the case and check whether our data are compatible with this. We choose a fixed value of the aspect ratio $L/\beta$,
corresponding to $z=1$,  so that the finite-size scaling function of these quantities 
involves only one variable, i.e. $O=  L^{-x_O} \tilde{O}(\delta\,L^{1/\nu})$. For a given $V_0$,
the deviation from the critical point is parameterized as $\delta=J-J_c$.  
The scaling dimension is given by $x_m=-\beta_m/\nu$ for the magnetization $|m|$,
$x_\chi=\gamma/\nu$ for the magnetic susceptibility $\chi$ and $x_g=0$ for
the dimensionless fourth cumulant $g$. Typically we executed $10^6-10^7$ MC steps to thermalize the system. 
Once in equilibrium, we generated $4-5\times10^7$ MC configurations  for measurements, 
which were carried out every 5 MC steps. We have considered a wide rage of values of
$V_0$ between $0.01$ and $5$.
At a fixed value of $V_0$ we varied the strength of the ferromagnetic interaction $J$ 
for system sizes up to $L=64$ to localize the zero-temperature critical point $J_c$ and to carry out 
the finite-size analysis.

\begin{figure}[!t]
 {\par\centering \resizebox*{0.9\textwidth}{!}
    {\includegraphics{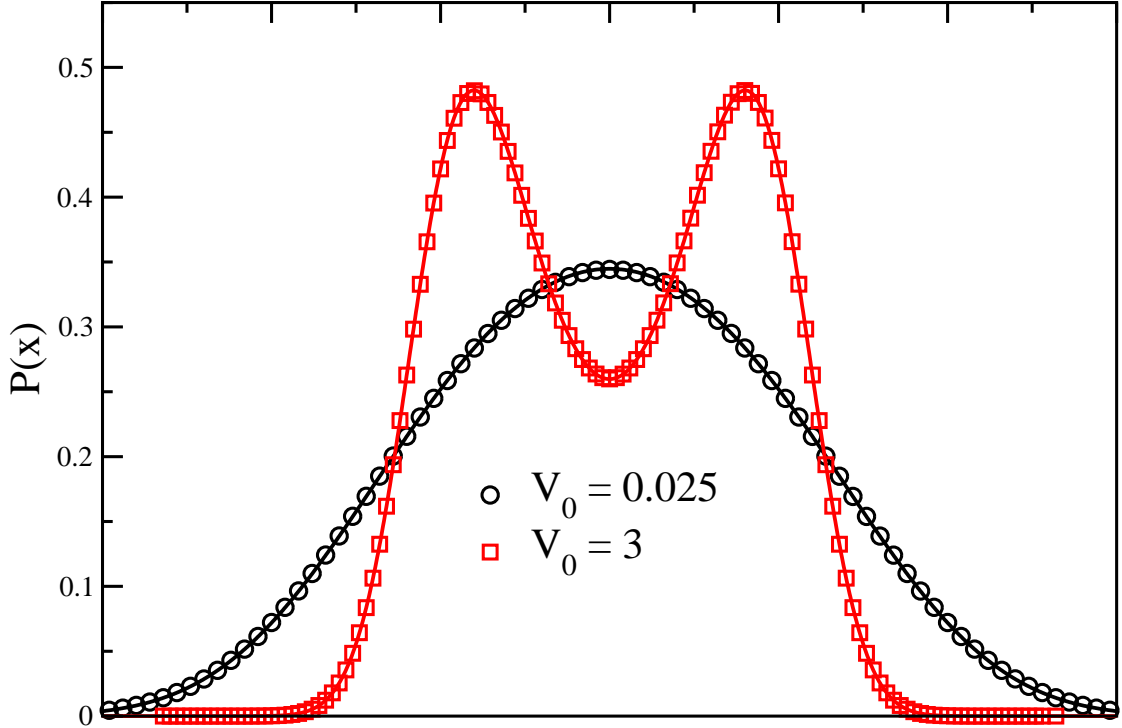}} \par}
 \caption{
 \label{fig:p(x)}
  The distribution of displacements of the particles, calculated for $L=32$ and $\beta=16$, 
  with $J=0$. A comparison with numerical solutions (indicated by the solid line) of 
  the Schr\"odinger equation is shown. The excellent agreement confirms the accuracy of the 
  density matrices.
 }
\end{figure}

\begin{figure}[!h]
 {\par\centering \resizebox*{0.9\textwidth}{!}
    {\includegraphics{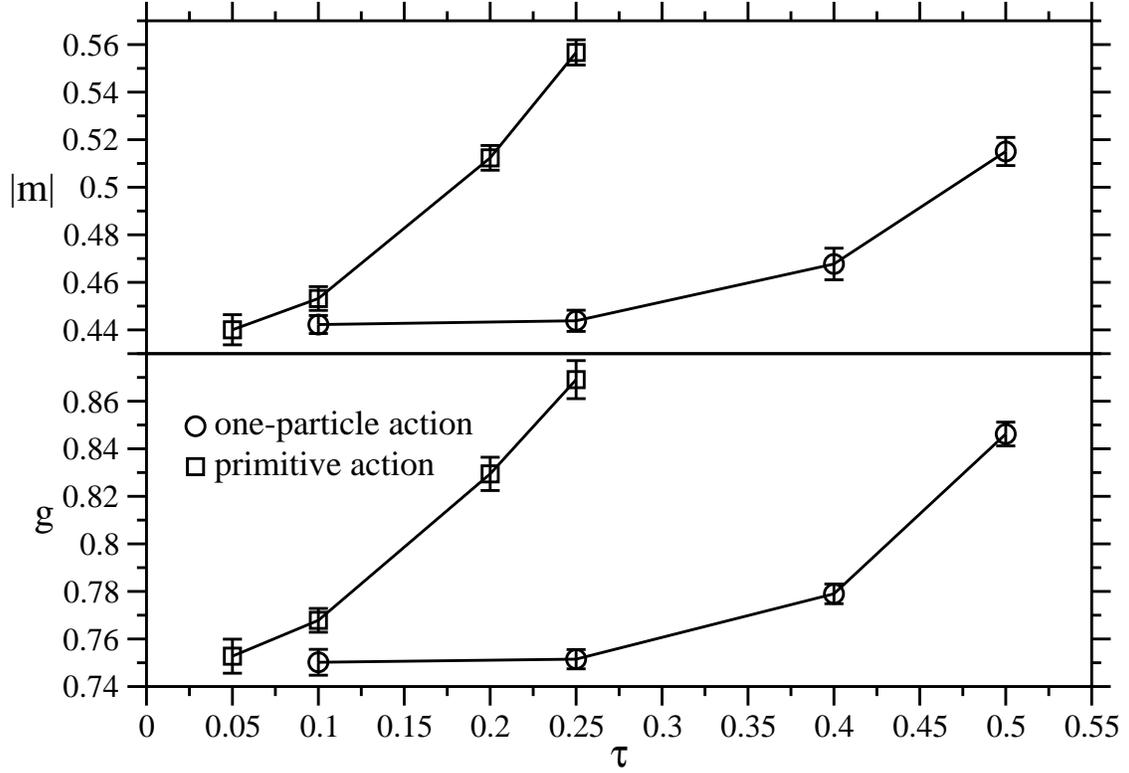}} \par}
 \caption{
 \label{fig:compare}
  The magnetization $\langle |m|\rangle$ and its fourth-order cumulant $g$,
  calculated by using the one-particle density matrix and the primitive approximation.
  The model parameters are $V_0=1$ and $J=0.56$ for a system size $L=32$ at temperature 
  $\beta=16$. The values obtained from both methods are compatible in the small $\tau$ limit.
 }
\end{figure}

To confirm the accuracy of the one-particle density matrices calculated by 
matrix multiplication method, we first compare the distribution of displacements of the particles 
in the absence of the interaction obtained by PIMC with the distribution calculated 
by solving numerically the single particle Schr\"odinger equations. For the latter, we calculated
the first $N=50$ energy eigenvalues $E_n$ and the corresponding eigenstates $\psi_n(x)$; 
the distribution function of the displacement is then computed by 
$P(x)=\sum_{n=1}^N |\psi_n(x)|^2\,e^{-\beta E_n}/\sum_{n=1}^N e^{-\beta E_n}$.   
As shown in Fig.~\ref{fig:p(x)} for $\beta=16$ by using $\tau=0.25$, the excellent 
agreement confirms the accuracy of the density matrices. Furthermore, we compare the results from PIMC within
the one-particle density matrix approximation with those in the primitive approximation for the same 
parameters (e.g. $V_0$ and $J$), as shown in Fig~\ref{fig:compare} as a typical example for the 
dependence of magnetization $|m|$ and its fourth-order cumulant $g$ on the time step $\tau$. 
The results presented are averaged over 16 samples for each time step. We find that, with $\tau=0.05$, 
the results obtained by the primitive approximation converge to the same values computed with the 
one-particle density matrix for $\tau\le 0.25$ within the statistical error bars. The CPU time required on 
an Intel Pentium processor (2.40GHz) to calculate 500000 MC steps for a system size $L=32$ and $\beta=16$ 
within the one-particle density matrix approximation using $\tau=0.25$ is about 780 seconds, and
with the primitive approximation using $\tau=0.05$ is about 1485 seconds. The efficiency of the
calculations with the one-particle density matrix is gained from the fast convergence with respect to
the number of time slice. After a careful examination, we are convinced that the time step $\tau=0.25$, used in
our simulations for the high-temperature one-particle density matrix, is sufficiently small for the 
convergence. We carried out $8$ iterations for the matrix multiplication to generate 
a one-particle density matrix with $\tau=0.25$ and the spacing between neighboring points within 
$[x_\textrm{min},x_\textrm{max}]$ was chosen to be $0.01$. In all cases studied we
used a wide interval of $[x_\textrm{min},x_\textrm{max}]$, e.g. $[-10,+10]$ 
for $V_0=3$, for the iterative integrations and then truncated this interval to a smaller one 
while storing into the look-up table for PIMC simulations. The appropriate values for the
interval $[x_\textrm{min},x_\textrm{max}]$ in the look-up table were justified by doing 
a short run of the PIMC simulation to check whether the particles would move beyond the chosen 
boundaries.

\begin{figure}[!h]
 {\par\centering \resizebox*{0.9\textwidth}{!}
    {\includegraphics{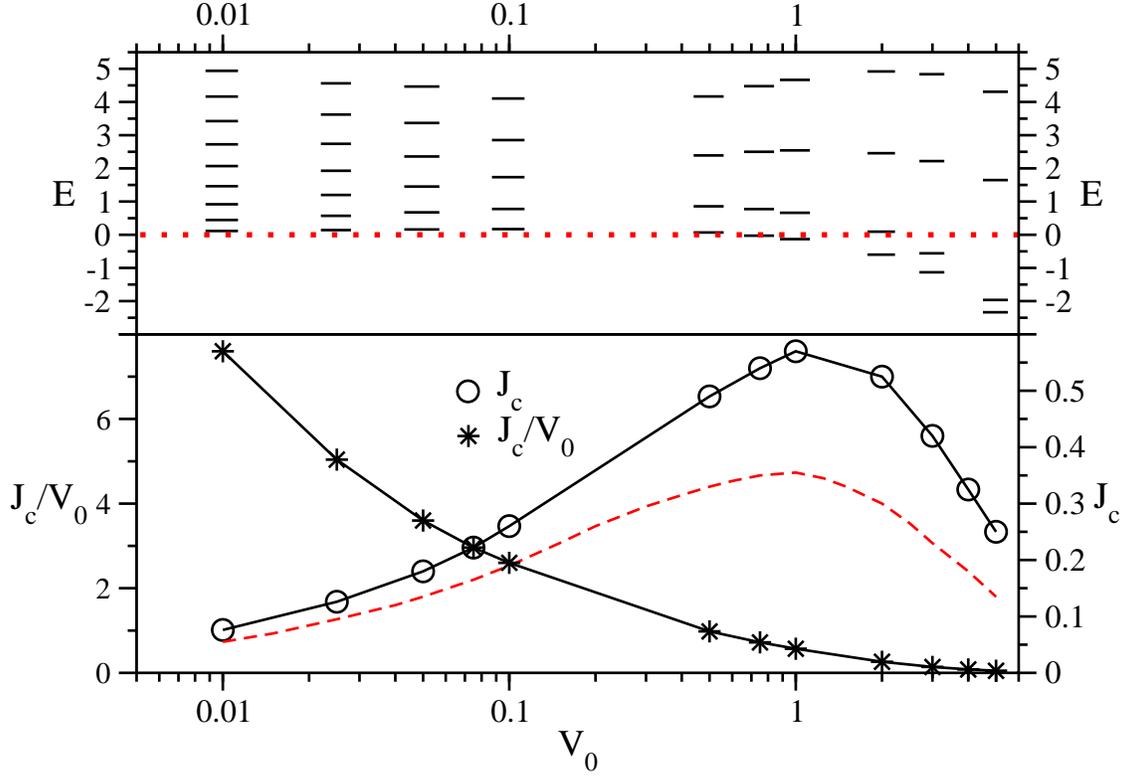}} \par}
 \caption{
 \label{fig:phase}
  Lower panel:
  The phase diagram of the coupled double-well chain: the critical ratio 
  $J_c/V_0$ as well as the critical interaction $J_c$ as functions of the depth of 
  the potential well $V_0$; the ordered phase is located above the curves and the 
  disordered phase is below the curves. The critical $J_c$ obtained by the
  mean field approach is indicated by the dashed line.
  Upper panel: The low lying energy eigenvalues of the one-particle Hamiltonian for 
  various $V_0$, determined by numerical solutions of the Schr\"odinger equation. 
  The dotted line at $E=0$ indicates the top of the potential barrier.
 }
\end{figure}

In Fig.~\ref{fig:phase} we present the zero-temperature phase diagram, 
in which the critical value $J_c$ is estimated by the intersection of $g(J)$ curves at a given $V_0$ 
for various system sizes (up to $L=64$) with fixed aspect ratio $L/\beta=2$. 
We note the lack of monotonicity of the critical $J_c$ with respect to the potential barrier $V_0$;
$J_c$ decreases  with $V_0$  in the deep well region,   
while it increases with $V_0$ in the small $V_0$ region.
This non-monotonic behavior of $J_c(V_0)$ is qualitatively reproduced within the  mean field 
approximation: Consider the effective single-site Hamiltonian including a mean-field term
\be
\mathcal{H}_{\textrm{mf}} = \frac{p^2}{2} + V_0(x^4-2x^2) - 2Jxm,
\ee
where the order parameter $m$ is the expectation value of the 
displacement $x$ in the ground state $\psi_0(x,m)$ of $\mathcal{H}_{\textrm{mf}}$ and is determined
self-consistently via
\be
 m=\int dx\,x |\psi_0(x,m)|^2.
 \label{eq:m_mf} 
\ee
Varying $J$ and solving the non-linear equation (\ref{eq:m_mf}) for $m$ numerically, 
the critical point can be estimated as the value of $J$ above which a non-zero solution exists.  

The mean-field result for $J_c$, depicted in Fig.~\ref{fig:phase}, shows the same non-monotonic 
behavior as our results for $J_c$ from the PIMC and has a maximum at $V_0=1$. This 
behavior of $J_c$ can be understood as follows: In the region $V_0 \gg 1$ the potential has
two deep minima separated by a barrier $V_0$ giving rise to a nearly degenerated ground state
doublet that is well separated from the rest of the spectrum, as shown in the upper panel of
Fig.~\ref{fig:phase}. This is called the order-disorder
limit\cite{BRUCE,RUBTSOV} in the interacting double-well model. The energy difference between the
ground state and the first excited state, i.e. the tunneling splitting $\Gamma$, is reduced as $V_0$ 
grows, which results in a decrease of the critical ordering term $J_c$. In the region $V_0 \ll 1$,
on the other hand, the potential has two shallow minima and the two lowest energy levels 
are not well separated from the rest of the spectrum. This is called the displacive 
regime\cite{NOTE} in the interacting double-well model. In this displacive region, 
the zero point energy of a single particle lies above the barrier of the local potential 
so that the local potential is effectively in a single-well form. Without the 
interaction term, the particles fluctuate around the $x=0$ position (cf. Fig.~\ref{fig:p(x)}); 
switching on the interaction shifts the displacement expectation value $\langle x \rangle$ away
from the origin and at the critical coupling $J_c$ the systems undergoes a displacive transition 
from a symmetric (disordered) phase to a broken symmetry (ordered) phase. The key factor
for the strength of the critical displacing force $J_c$ in this case is the width of 
the local potential, which decreases with increasing $V_0$: the wider the local potential,
the weaker the force $J$ needed for the displacement. Therefore, the critical value
$J_c$ increases with $V_0$  in the displacive regime.

\begin{figure}[h!]
 {\par\centering \resizebox*{0.9\textwidth}{!}
    {\includegraphics{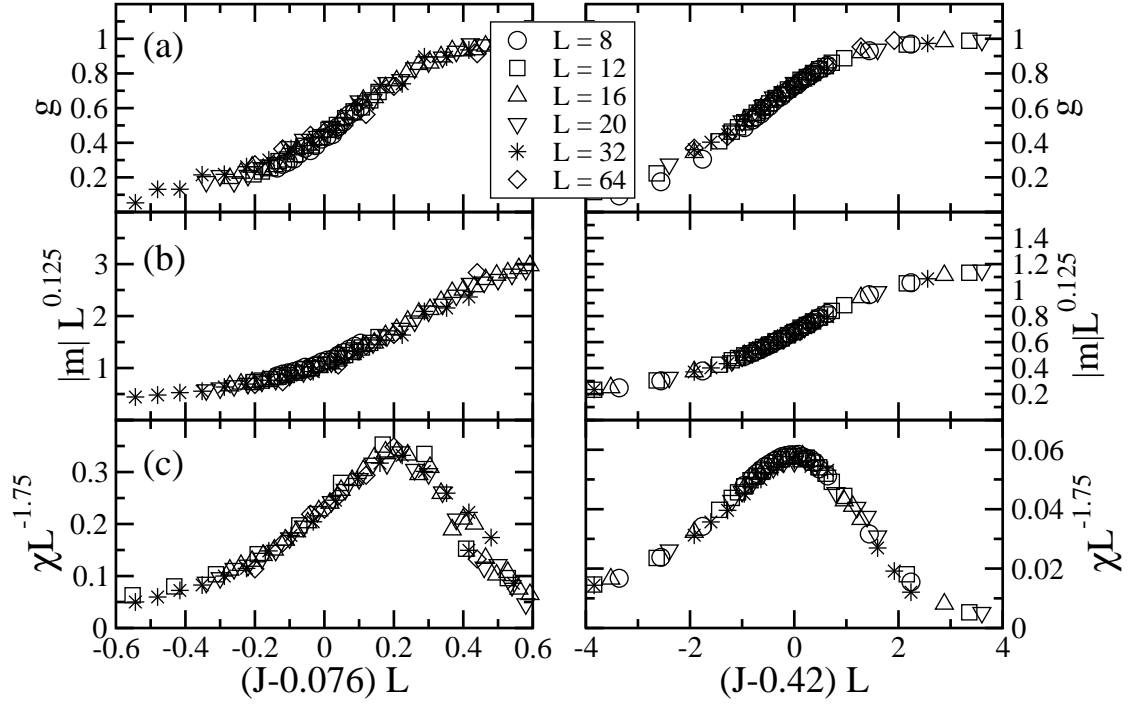}} \par}
 \caption{
 \label{fig:scaling}
  Scaling plots of the cumulant(a), the magnetization(b) and 
  the susceptibility(c) for $V_0=0.01$ (left) and $V_0=3$ (right) using
  the two-dimensional Ising universality.
 }
\end{figure}

\begin{figure}[h!]
 {\par\centering \resizebox*{0.9\textwidth}{!}
    {\includegraphics{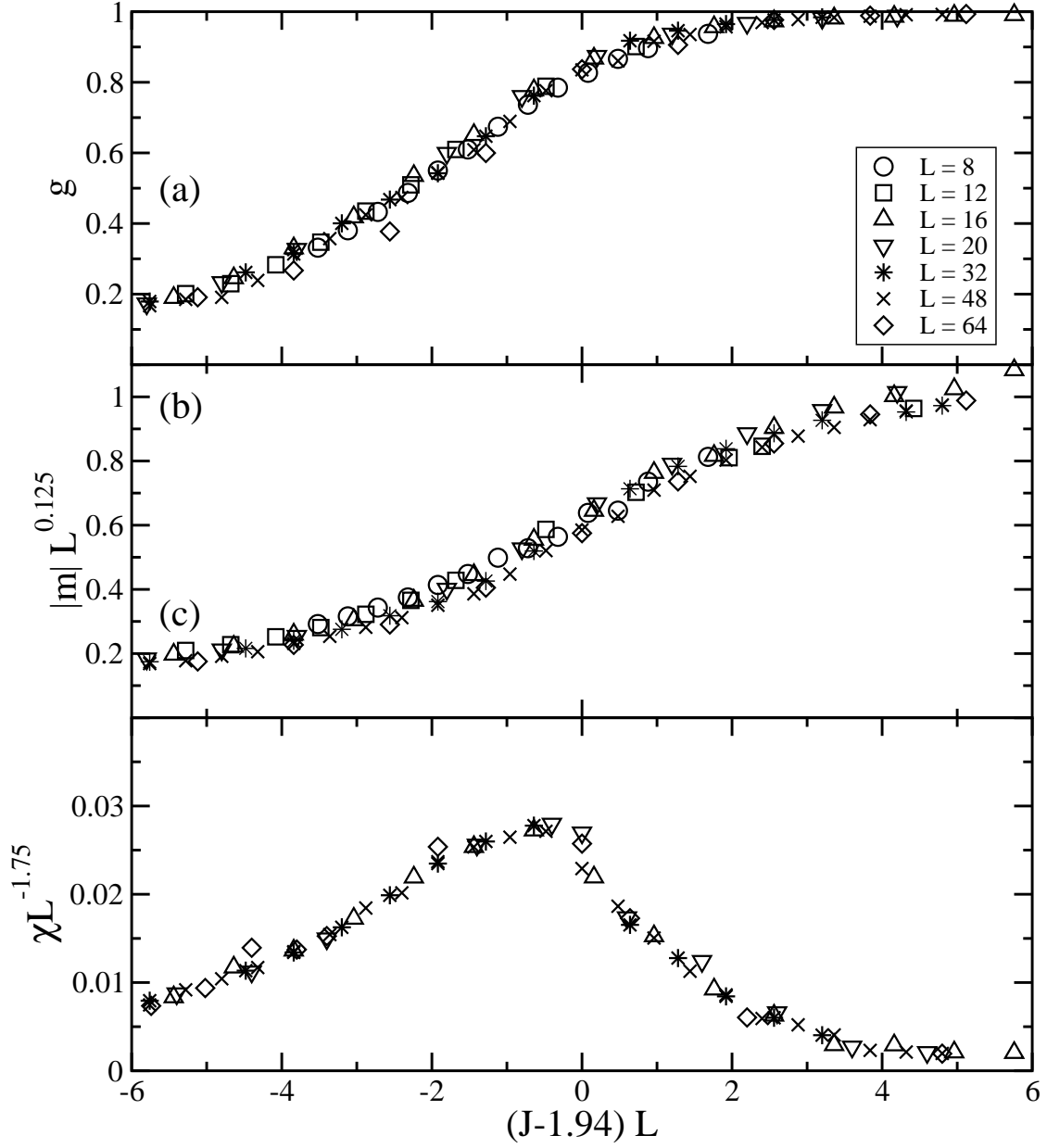}} \par}
 \caption{
 \label{fig:x4}
  Scaling plots of the cumulant(a), the magnetization(b) and the susceptibility(c) 
  for the single-well model with $U(x)=x^4$ using the two-dimensional Ising universality. 
  In the scaling plot (c), the system sizes only range from $L=16$ to $L=64$.
 }
\end{figure}

For a particular value of $V_0$ we can use the scaling form given in Eq.~(\ref{eqn:scaling}) 
for $g$, $|m|$ and $\chi$ to extract values of the critical exponents. In all cases a good
data collapse is achieved with the exponents $\beta_m=1/8,\,\gamma=7/4$ and $\nu=1.0$, 
which is representative of the classical 2D Ising universality class. In Fig.~\ref{fig:scaling}
we show the finite-size scaling plots for $V_0=3$ and $V_0=0.01$. For $V_0=0.01$ which is well
inside the displacive regime, the quality of the scaling decreases and corrections to
scaling become more pronounced. Interestingly, the peak of the scaling function $\tilde{\chi}(t)$
of the susceptibility is shifted away from $t=0$ for $V_0=0.01$, whereas it is at $t=0$ 
for $V_0=3$, indicating the non-universality of the scaling function. Our results for
$V_0 \ll 1$ indicate that the model is in the Ising universality class even in the
displacive regime. For the interaction in the form given in Eq.~(\ref{VV}), we expect that
a phase transition in the same universality still occurs when the local potential
is reduced to only a quartic term. To provide support for this we
carried out simulations for the model with a local potential given by
$U(x)=x^4$ which exhibits a single well. Our results depicted in Fig.~\ref{fig:x4} suggest
that the critical behavior of this one-well model is indeed consistent with 2D Ising
universality\cite{BOADER}. We note that the coupling term (\ref{VV}) that we use can be 
brought into a form that is more reminiscent of a lattice version of the standard
$\phi^4$ (quantum) field theory:
\be
{\cal H}_B=\frac{J}{2}\sum_{i=1}^L (x_i-x_{i+1})^2
- J\sum_{i=1}^L x_i^2\;.
\label{form}
\ee
Together with the local potential (\ref{UU}) this implies that the
corresponding continuum model for a scalar field $\phi$ contains a
$(d\phi/dx)^2$-term and a potential energy of the form
$V_0[\phi^4-(2+J/V_0)\phi^2)]$. Since $J$ is always positive and can
be made arbitrarily large, this model has always a phase transition
(at zero temperature). On the other hand, the field theory with a
potential energy that has only a single minimum, like the pure quartic
potential $V_0\phi^4$, does {\it not} have a phase transition. We
checked, within mean-field as well as with PIMC simulations, that the
corresponding lattice model
\be
{\cal H}=\sum_{i=1}^L \left\{\frac{p_i^2}{2m}
+ \frac{J}{2}(x_i-x_{i+1})^2 + V_0 x_i^4\right\}
\ee
also does not have a phase transition.

To summarize, we have demonstrated that PIMC within the one-particle density matrix 
approximation is an efficient method to simulate quantum interacting many-body systems, 
in which particles are confined in a local potential and interact
with each other. Using this method we have studied the zero-temperature
phase transition of the coupled double-well chain, both in the order-disorder case, 
corresponding to a coupled two-level tunneling system, and in the displacive regime, 
in which the interaction dominates over the double-well structure.
Based on this numerical scheme, our further study will include the double-well model 
coupled through long-range/random interactions and coupled to a dissipative bath\cite{VAIA,KATO}. 
In the presence of quenched disorder in the coupling, even for the case without dissipation, 
implementation of many improved PIMC methods, e.g. Fourier PIMC techniques or cluster 
algorithms, becomes complex and the computational efficiency reduces. 
This motivates the choice of a method which provides easy performance and can
be extended to the random case in a straightforward way, as the technique
applied in this paper does.

\clearpage

%%%%%%%%%%%%%%%%%%%%%%%%%%%%% Fig 1 %%%%%%%%%%%%%%%%%%%%%%%%%%%%%%%%%%%%%%%

\clearpage
%%%%%%%%%%%%%%%%%%%%%%%%%%%%%%%%%%%%%%%%%%%%%%%%%%%%%%%%%%%%%%%%%%%%%%%%%%

%%%%%%%%%%%%%%%%%%%%%%%%%%%%% Fig 2 %%%%%%%%%%%%%%%%%%%%%%%%%%%%%%%%%%%%%%%

\clearpage
%%%%%%%%%%%%%%%%%%%%%%%%%%%%%%%%%%%%%%%%%%%%%%%%%%%%%%%%%%%%%%%%%%%%%%%%%%

%%%%%%%%%%%%%%%%%%%%%%%%%%%%%%% Fig 3 %%%%%%%%%%%%%%%%%%%%%%%%%%%%%%%%%%%%%

\clearpage

%%%%%%%%%%%%%%%%%%%%%%%%%%%%%%%%%%%%%%%%%%%%%%%%%%%%%%%%%%%%%%%%%%%%%%%%%%%%

%%%%%%%%%%%%%%%%%%%%%%%%%%%%%%% Fig 4 %%%%%%%%%%%%%%%%%%%%%%%%%%%%%%%%%%%%%%%

\clearpage
%%%%%%%%%%%%%%%%%%%%%%%%%%%%%%%%%%%%%%%%%%%%%%%%%%%%%%%%%%%%%%%%%%%%%%%%%%%%%

%%%%%%%%%%%%%%%%%%%%%%%%%%%%%%% Fig 5 %%%%%%%%%%%%%%%%%%%%%%%%%%%%%%%%%%%%%%%

\clearpage
%%%%%%%%%%%%%%%%%%%%%%%%%%%%%%%%%%%%%%%%%%%%%%%%%%%%%%%%%%%%%%%%%%%%%%%%%%%%%%%%

%%%%%%%%%%%%%%%%%%%%%%%%%%%%%%% Fig 6 %%%%%%%%%%%%%%%%%%%%%%%%%%%%%%%%%%%%%%%%%%

\clearpage
%%%%%%%%%%%%%%%%%%%%%%%%%%%%%%%%%%%%%%%%%%%%%%%%%%%%%%%%%%%%%%%%%%%%%%%%%%%%%%%

%%%%%%%%%%%%%%%%%%%%%%%%%%%%%%% Fig 7 %%%%%%%%%%%%%%%%%%%%%%%%%%%%%%%%%%%%%%%%

%%%%%%%%%%%%%%%%%%%%%%%%%%%%%%%%%%%%%%%%%%%%%%%%%%%%%%%%%%%%%%%%%%%%%%%%%%%%%%%%%


\begin{thebibliography}{99}
\bibitem{NOTE} In some literature on the classical $\phi^4$ model the displacive region is defined 
               as the region where the local potential exhibits a single well 
               (see e.g. \cite{SCHNEIDER,BINDER}). 
               In this manuscript we use the term for the low-barrier region in our quantum model.
\end{thebibliography}
\end{document}